\documentclass[journal]{IEEEtran}

\usepackage{graphicx}
\usepackage{color}
\usepackage{placeins}
\usepackage{float}
\usepackage{tabularx,colortbl}
\usepackage{amssymb}
\usepackage{amsthm}
\usepackage{cite}
\usepackage{amsmath}

\theoremstyle{plain}

\newtheorem*{lemm*}{Lemma}
\theoremstyle{plain}

\newtheorem{remark}{Remark}

\IEEEoverridecommandlockouts

\begin{document}

\title{Uplink Performance of Cell-Free Massive MIMO Over Spatially Correlated Rician Fading Channels
\thanks{Manuscript received xxxxxxxxxxxxxxxx; revised xxxxxxxxxxxxxxxx; accepted November xxxxxxx. Date of publication xxxxxxxxxxxxxxxxx; date of
current version xxxxxxxxxxxxxxx. The associate editor coordinating the review of this letter and approving it for publication was xxxxxx. (Corresponding author: Jiayi Zhang.)}
\thanks{Z. Wang and J. Zhang are with the School of Electronic and Information Engineering, Beijing Jiaotong University, Beijing 100044, China.(e-mail: \{20111050, jiayizhang\}@bjtu.edu.cn).}
\thanks{E. Bj{\"o}rnson  is  with  the  KTH  Royal  Institute  of  Technology, 10044 Stockholm, Sweden, and Link{\"o}ping University,  58183  Link{\"o}ping, Sweden (e-mail: emilbjo@kth.se)}
\thanks{B. Ai is with the State Key Laboratory of Rail Traffic Control and Safety, Beijing Jiaotong University, Beijing 100044, China (e-mail: aibo@ieee.org).}}
\author{Zhe Wang, Jiayi Zhang,~\IEEEmembership{Senior Member,~IEEE,} \\Emil Bj{\"o}rnson,~\IEEEmembership{Senior Member,~IEEE,} and Bo Ai,~\IEEEmembership{Senior Member,~IEEE}}
\maketitle

\begin{abstract}
We consider a practical cell-free massive multiple-input-multiple-output (MIMO) system with multi-antenna access points (APs) and spatially correlated Rician fading channels. The significant phase-shift of the line-of-sight component induced by the user equipment movement is modeled randomly. Furthermore, we investigate the uplink spectral efficiency (SE) with maximum ratio (MR)/local minimum mean squared error (L-MMSE) combining and optimal large-scale fading decoding based on the phase-aware MMSE, phase-aware element-wise MMSE and linear MMSE (LMMSE) estimators. Then new closed-form SE expressions with MR combining are derived. Numerical results validate our derived expressions and show that the SE benefits from the spatial correlation. It is important to observe that the performance gap between L-MMSE and MR combining increases with the number of antennas per AP and the SE of the LMMSE estimator is lower than that of other estimators due to the lack of phase-shifts knowledge.
\end{abstract}

\begin{IEEEkeywords}
cell-free massive MIMO, spatially correlated Rician fading, phase-shift, spectral efficiency.
\end{IEEEkeywords}

\IEEEpeerreviewmaketitle

\section{Introduction}
{As one of the most promising technologies for future wireless communication, cell-free massive multiple-input-multiple-output (CF mMIMO) has been widely investigated in \cite{7827017,zhang2019multiple,7917284,8901451,8943119}. The key concept is that a large number of access points (APs) are connected to the central processing unit (CPU) via fronthaul connections to jointly serve the user equipments (UEs) on the same time-frequency resource. The number of APs is envisioned to be much larger than the number of UEs, thus distances between the closest AP-UE pairs decrease greatly, which leads to the decrease in path loss and the increase in macro diversity gain. In the uplink (UL), CF mMIMO usually uses maximum ratio (MR) combining for the low complexity but \cite{8845768} advocates for local minimum mean squared error (L-MMSE) combining for its better performance. Moreover, the large-scale fading decoding (LSFD) method proposed for mMIMO originally has been utilized in CF mMIMO systems to further improve the throughput \cite{8845768}.

{The vast majority of scientific papers on CF mMIMO are making the simplifying assumption of Rayleigh fading channels \cite{8845768,9004558,8891922,8886730} or Rician fading channels where the line-of-sight (LoS) component has a static phase \cite{8869794}. Recently, the authors in \cite{8809413} indicate that the practical channel in CF mMIMO should be composed of a semi-deterministic LoS path component with random phase-shifts and a stochastic non-line-of-sight (NLoS) path component. The phase-shift of the LoS component is modeled as a uniformly distributed random variable due to UE mobility and hardware effects like phase noise. However, \cite{8809413} is based on the assumption of single-antenna APs, while practical APs are usually equipped with multiple antennas. Furthermore, the authors in \cite{133} consider spatially uncorrelated Rician fading channels with unknown phase shifts and multi-antenna APs. Unfortunately, it did not consider the spatial channel correlation which has a significant impact on CF mMIMO systems \cite{8620255}.}

To address these limitations, we consider a CF mMIMO system over spatially correlated Rician fading channels with phase-shifts and multi-antenna APs. The same channel model has been investigated in the cellular mMIMO scenario in \cite{9053886}. The main contributions of this paper are as follows: (1) We consider three useful channel estimators with different prior information: the phase-aware MMSE with all prior information, the phase-aware element-wise MMSE (EW-MMSE) with phase-shifts and partial large-scale fading knowledge and the linear MMSE (LMMSE) with all large-scale fading but no phase-shift knowledge; (2) Based on these channel estimators and the LSFD method, we derive the UL SE expressions for any combining scheme and compute closed-form SE expressions for MR combining; (3) We analyse the UL SE with MR/L-MMSE combining over correlated/uncorrelated Rician fading channels numerically.

\section{System Model}\label{se:model}
We consider a CF mMIMO system consisting of $M$ APs with $N$ antennas each and $K$ single-antenna UEs. The channel response is constant in a coherence time-frequency block of length $\tau _c$ channel uses. In the UL, we reserve $\tau _p$ channel uses for the training and $\tau _u=\tau _c-\tau _p$ channel uses for the data transmission. Let $\mathbf{h}_{mk}\in \mathbb{C}^N$ denote the channel between AP $m$ and UE $k$. We assume $\mathbf{h}_{mk}$ is an independent random variable for every AP $m$-UE $k$ pair and $\mathbf{h}_{mk}$ in different coherence blocks are independent and identically distributed (i.i.d.). We consider the spatially correlated Rician fading channel which is composed of a semi-deterministic LoS path component and a stochastic NLoS path component as
 \begin{equation}\label{eq:er_model}
  \mathbf{h}_{mk}=\mathbf{\Phi }_{mk}\mathbf{\bar{h}}_{mk}+\mathbf{g}_{mk},
\end{equation}
where $\mathbf{g}_{mk}\sim \mathcal{N}_{\mathbb{C}}( \mathbf{0},\mathbf{R}_{mk} )$ is the NLoS component and $\mathbf{R}_{mk}\in \mathbb{C}^{N\times N}$ is the spatial correlation matrix. $\beta _{mk}^{\text{NLoS}}=\frac{\text{tr}\left( \mathbf{R}_{mk} \right)}{N}$ denotes the large-scale fading coefficient for the NLoS propagation. $\mathbf{\bar{h}}_{m,k}\in \mathbb{C}^N$ represents the deterministic LoS component. Moreover, $\mathbf{\Phi }_{mk}=\text{diag}( e^{j\varphi _{mk1}},\cdots ,e^{j\varphi _{mkN}} ) \in \mathbb{C}^{N\times N}$ where $\varphi _{mkn}\sim \mathcal{U}\left[ -\pi ,\pi \right]$ is the additional phase-shift of the LoS component between the $n$-th antenna of AP $m$ and UE $k$.  In this paper, we assume all elements of $\mathbf{\Phi }_{m,k}$ are equal, so the LoS component in \eqref{eq:er_model} can be denoted as $\mathbf{\bar{h}}_{mk}e^{j\varphi _{mk}}$. Note that $\varphi _{mk}\sim \mathcal{U}\left[ -\pi ,\pi \right]$ vary at the same pace as $\mathbf{g}_{mk}$ and $\varphi _{mk}$ in different coherence blocks are assumed to be i.i.d.
\begin{remark}
We notice that \eqref{eq:er_model} is a multi-antenna generalization of \cite{8809413} and an extension of \cite{133} to spatially correlated Rician fading channels.
\end{remark}
\begin{remark}
We can treat each $N$-antenna AP as a cluster of $N$ single-antenna APs only if the channel coefficients to the $N$ antennas of an AP are independently distributed.
\end{remark}

\subsection{Channel Estimation}
We use $\tau _p$ mutually orthogonal pilot sequences for channel estimation. $\boldsymbol{\phi }_k\in \mathbb{C}^{\tau _p}$ denotes the pilot sequence of UE $k$, with $\lVert \boldsymbol{\phi }_k \rVert ^2=\tau _p$. Notice that $K>\tau _p$, so more than one UE use the same pilot sequence. We define $\mathcal{P}_k$ as the index subset of UEs that use the same pilot sequence as UE $k$ including itself. The received signal $\mathbf{y}_{m}^{p}\in \mathbb{C}^{N\times \tau _p}$ at AP $m$ is given by
\begin{equation}
\mathbf{y}_{m}^{p}=\sum_{k=1}^K{\sqrt{\hat{p}_k}\mathbf{h}_{mk}\boldsymbol{\phi }_{k}^{T}+\mathbf{n}_{m}^{p}},
\end{equation}
where $\hat{p}_k$ is the pilot transmit power of UE $k$, $\mathbf{n}_{m}^{p}\in \mathbb{C}^{N\times \tau _p}$ is additive noise with independent $\mathcal{N}_{\mathbb{C}}( 0,\sigma ^2 ) $ entries, and $\sigma ^2$ is the noise power. In order to estimate $\mathbf{h}_{mk}$, AP $m$ multiplies $\mathbf{y}_{m}^{p}$ with pilot sequence of UE $k$ to obtain $\mathbf{y}_{mk}^{p}=\mathbf{y}_{m}^{p}\boldsymbol{\phi }_{k}^{*}$ as
\begin{equation}\label{eq:er_pilot}
\mathbf{y}_{mk}^{p}=\sqrt{\hat{p}_k}\tau _p\mathbf{h}_{mk}+\!\!\!\sum_{l\in \mathcal{P}_k\backslash\left\{ k\right\}}^K\!\!\!{\sqrt{\hat{p}_l}\mathbf{h}_{ml}\boldsymbol{\phi }_l\boldsymbol{\phi }_{k}^{*}+\mathbf{n}_{m}^{p}}\boldsymbol{\phi }_{k}^{*}.
\end{equation}

Based on \eqref{eq:er_pilot}, we can derive three useful channel estimators with different prior information. We will focus on the effects of phase-shifts and spatial correlation matrices in the following.
\subsubsection{Phase-Aware MMSE Estimator}
If $\mathbf{\bar{h}}_{mk}$, $\mathbf{R}_{mk}$ and $\varphi _{mk}$ are available for AP $m$, we can derive the phase-aware MMSE estimate of $\mathbf{h}_{mk}$ as
\begin{equation}\label{eq:mmse}
\mathbf{\hat{h}}_{mk}^{\text{mmse}}=\mathbf{\bar{h}}_{mk}e^{j\varphi _{mk}}+\sqrt{\hat{p}_k}\mathbf{R}_{mk}\mathbf{\Psi }_{mk}^{-1}\left( \mathbf{y}_{mk}^{p}-\mathbf{\bar{y}}_{mk}^{p} \right),
\end{equation}
where $\mathbf{\bar{y}}_{mk}^{p}=\sum\nolimits_{l\in \mathcal{P}_k}^{}{\sqrt{\hat{p}_l}\tau _p}\mathbf{\bar{h}}_{ml}e^{j\varphi _{ml}}$ and $\mathbf{\Psi }_{mk}=\sum\nolimits_{l\in \mathcal{P}_k}^{}{\hat{p}_l\tau _{p}^{}\mathbf{R}_{ml}}+\sigma ^2\mathbf{I}_N$. $\varphi _{mk}$, $\mathbf{y}_{mk}^{p}$ and $\mathbf{\bar{y}}_{mk}^{p}$ change in every coherence block so that \eqref{eq:mmse} is a single realization. The channel estimate $\mathbf{\hat{h}}_{mk}^{\text{mmse}}$ and estimation error $\mathbf{\tilde{h}}_{mk}^{\text{mmse}}=\mathbf{h}_{mk}-\mathbf{\hat{h}}_{mk}^{\text{mmse}}$ are independent random variables with
\begin{align}
&\mathbb{E}\left\{ \mathbf{\hat{h}}_{mk}^{\text{mmse}}\!\left| \varphi _{mk} \right. \right\}\!=\!\mathbf{\bar{h}}_{mk}e^{j\varphi _{mk}}\!, \text{Cov}\left\{ \mathbf{\hat{h}}_{mk}^{\text{mmse}}\!\left| \varphi _{mk} \right. \right\} \!=\!\hat{p}_k\tau _p\mathbf{\Omega }_{mk},\notag\\
&\mathbb{E}\left\{ \mathbf{\tilde{h}}_{mk}^{\text{mmse}}\right \} =\mathbf{0},\qquad\quad\quad \ \ \ \ \text{Cov}\left\{ \mathbf{\tilde{h}}_{mk}^{\text{mmse}} \right\} =\mathbf{C}_{mk}^{\text{mmse}},\notag
\end{align}
where $\mathbf{\Omega }_{mk}=\mathbf{R}_{mk}\mathbf{\Psi }_{mk}^{-1}\mathbf{R}_{mk}$ and $\mathbf{C}_{mk}^{\text{mmse}}=\mathbf{R}_{mk}-\hat{p}_k\tau _p\mathbf{R}_{mk}\mathbf{\Psi }_{mk}^{-1}\mathbf{R}_{mk}$.
\subsubsection{Phase-Aware EW-MMSE Estimator}
If $\mathbf{\bar{h}}_{mk}$, $\varphi _{mk}$ and the diagonals of $\mathbf{R}_{mk}$ are available for AP $m$, we can obtain the phase-aware EW-MMSE estimation of $\mathbf{h}_{mk}$ as
\begin{equation}
\mathbf{\hat{h}}_{mk}^{\text{ew}}=\mathbf{\bar{h}}_{mk}e^{j\varphi _{mk}}+\sqrt{\hat{p}_k}\mathbf{D}_{mk}\mathbf{\Lambda }_{mk}^{-1}\left( \mathbf{y}_{mk}^{p}-\mathbf{\bar{y}}_{mk}^{p} \right),
\end{equation}
where $\mathbf{D}_{mk}\triangleq\text{diag}([\mathbf{R}_{mk}]_{nn}:n=1,\cdots,N)$ and $\mathbf{\Lambda}_{mk}\triangleq\text{diag}([\mathbf{\Psi}_{mk}] _{nn}:n=1,\cdots,N)$. The channel estimate $\mathbf{\hat{h}}_{mk}^{\text{ew}}$ and estimation error $\mathbf{\tilde{h}}_{mk}^{\text{ew}}=\mathbf{h}_{mk}-\mathbf{\hat{h}}_{mk}^{\text{ew}}$ are correlated random variables with
\begin{align}
&\mathbb{E}\left\{ \mathbf{\hat{h}}_{mk}^{\text{ew}}\left| \varphi _{mk} \right. \right\} =\mathbf{\bar{h}}_{mk}e^{j\varphi _{mk}},\text{Cov}\left\{ \mathbf{\hat{h}}_{mk}^{\text{ew}}\left| \varphi _{mk} \right.  \right\} =\mathbf{\Sigma }_{mk},\notag\\
&\mathbb{E}\left\{ \mathbf{\tilde{h}}_{mk}^{\text{ew}} \right\} =\mathbf{0},\qquad\qquad\quad\ \ \ \text{Cov}\left\{ \mathbf{\tilde{h}}_{mk}^{\text{ew}} \right\} =\mathbf{C}_{mk}^{\text{ew}},\notag
\end{align}
where $\mathbf{\Sigma }_{mk} \triangleq \hat{p}_k\tau _p\mathbf{D}_{mk}\mathbf{\Lambda }_{mk}^{-1}\mathbf{\Psi }_{mk}\mathbf{\Lambda }_{mk}^{-1}\mathbf{D}_{mk}$ and $\mathbf{C}_{mk}^{\mathrm{ew}}\!\!\triangleq\!\mathbf{R}_{mk}-\hat{p}_k\tau _p( \mathbf{R}_{mk}\mathbf{\Lambda }_{mk}^{-1}\mathbf{D}_{mk}-\mathbf{D}_{mk}\mathbf{\Lambda }_{mk}^{-1}\mathbf{R}_{mk} ) +\mathbf{\Sigma }_{mk}$.
\subsubsection{LMMSE Estimator}
If $\mathbf{\bar{h}}_{mk}$ and $\mathbf{R}_{mk}$ are available and the phase-shift $\varphi_{mk}$ is unknown at AP $m$, the LMMSE estimate of $\mathbf{h}_{mk}$ is
\begin{equation}
\mathbf{\hat{h}}_{mk}^{\text{lmmse}}=\sqrt{\hat{p}_k}\mathbf{R}'_{mk}\left( \mathbf{\Psi }'_{mk} \right) ^{-1}\mathbf{y}_{mk}^{p},
\end{equation}
where $\mathbf{R}'_{mk}\!\!\!\!\!\!\!\!\triangleq\!\!\!\!\!\!\mathbf{R}_{mk}+\mathbf{\bar{h}}_{mk}\mathbf{\bar{h}}_{mk}^{H}$ and $\mathbf{\Psi}'_{mk}\!\!\!\!\triangleq\!\sum\nolimits_{l\in \mathcal{P}_k}^{}\!{\begin{array}{c}\!\hat{p}_l\tau _p\mathbf{R}'_{ml}+\sigma ^2\mathbf{I}_N\\\end{array}}$. The channel estimate $\mathbf{\hat{h}}_{mk}^{\text{lmmse}}$ and estimation error $\mathbf{\tilde{h}}_{mk}^{\text{lmmse}}\!=\!\mathbf{h}_{mk}-\mathbf{\hat{h}}_{mk}^{\text{lmmse}}$ are uncorrelated random variables with
\begin{align}
&\mathbb{E}\left\{ \mathbf{\hat{h}}_{mk}^{\text{lmmse}} \right\} =\mathbf{0},\text{Cov}\left\{ \mathbf{\hat{h}}_{mk}^{\text{lmmse}} \right\} =\hat{p}_k\tau _p\mathbf{\Omega }'_{mk},\notag\\
&\mathbb{E}\left\{ \mathbf{\tilde{h}}_{mk}^{\text{lmmse}}\right \} =\mathbf{0},\text{Cov}\left\{ \mathbf{\tilde{h}}_{mk}^{\text{lmmse}}\right \} =\mathbf{C}_{mk}^{\text{lmmse}},\notag
\end{align}
where $\mathbf{\Omega }'_{mk}\!\!=\!\!\mathbf{R}'_{mk}( \mathbf{\Psi }'_{mk} ) ^{-1}\mathbf{R}'_{mk}$ and $\mathbf{C}_{mk}^{\text{lmmse}}=\mathbf{R}'_{mk}-\hat{p}_k\tau _p\mathbf{R}'_{mk}( \mathbf{\Psi }'_{mk} ) ^{-1}\mathbf{R}'_{mk}$.

\subsection{UL Data Transmission}
In the UL, all UEs simultaneously send $\tau_u$ UL data symbols per coherence block to the APs. The received signal $\mathbf{y}_m\in \mathbb{C}^N$ at AP $m$ is
\begin{equation}
\mathbf{y}_m=\sum_{k=1}^K{\mathbf{h}_{mk}s_k}+\mathbf{n}_{m}^{\text{ul}},
\end{equation}
where $s_k\sim \mathcal{N}_{\mathbb{C}}(0,p_k )$ is the UL signal transmitted by UE $k$ with power $p_k=\mathbb{E}\{ | s_k |^2 \}$ and $\mathbf{n}_{m}^{\text{ul}}\sim \mathcal{N}_{\mathbb{C}}( 0,\sigma ^2\mathbf{I}_N )$ is the independent noise. Every AP can detect the UL data locally with a receive combining vector. Let $\mathbf{v}_{mk}$ denote the combining vector designed by AP $m$ for UE $k$ and the local estimate of $s_k$ in AP $m$ is given by
\begin{equation}\label{eq:sk}
\tilde{s}_{mk} \!=\!\mathbf{v}_{mk}^{H}\mathbf{h}_{mk}s_k+\!\!\!\sum_{l=1,l\ne
k}^K\!\!\!{\mathbf{v}_{mk}^{H}\mathbf{h}_{ml}s_l}+\mathbf{v}_{mk}^{H}\mathbf{n}_{m}^{\text{ul}}.
\end{equation}

Any combining vector is available for \eqref{eq:sk} and AP $m$ can use its local channel state information (CSI) to design $\mathbf{v}_{mk}$. We consider two combining schemes: MR combining with $\mathbf{v}_{mk}=\mathbf{\hat{h}}_{mk}^{i}$ where $i \in \{ \text{mmse},\text{ew},\text{lmmse} \}$ correspond to the MMSE, EW-MMSE and LMMSE estimators, respectively, and L-MMSE combining as
\begin{equation}\label{eq:L_MMSE}
\mathbf{v}_{mk}\!=\!p_k\left( \sum_{l=1}^K{p_l\left( \mathbf{\hat{h}}_{ml}^{i}\left( \mathbf{\hat{h}}_{ml}^{i}\right)^H\!\!\!\!+\mathbf{C}_{ml}^{i} \right)}\!\!+\sigma ^2\mathbf{I}_N \right)^{-1}\!\!\mathbf{\hat{h}}_{mk}^{i}.
\end{equation}
Note that \eqref{eq:L_MMSE} is optimal for the MMSE and LMMSE estimators since it can minimize $\text{MSE}_{mk}=\mathbb{E}\{ | s_k-\mathbf{v}_{mk}^{H}\mathbf{y}_m |^2 \left|\right.\{ \mathbf{\hat{h}}_{mk}^{i} \} \}$, but suboptimal for the EW-MMSE estimator.

To further mitigate the inter-user interference, the local estimates $\{ \tilde{s}_{mk}:m=1,\cdots ,M \}$ are sent to the CPU where they are linearly weighted by the LSFD coefficients to derive $\hat{s}_k=\sum\nolimits_{m=1}^M{\alpha _{mk}^{*}\tilde{s}_{mk}}$ as
\begin{equation}\label{eq:sk_LSFD}
\hat{s}_k=\mathbf{a}_{k}^{H}\mathbf{b}_{kk}s_k+\sum_{l=1,l\ne k}^K{\mathbf{a}_{k}^{H}\mathbf{b}_{kl}s_l}+\mathbf{n}_k,
\end{equation}
where $\mathbf{a}_k=[ \alpha _{1k},\cdots ,\alpha _{Mk} ] ^T\in \mathbb{C}^{M}$ is the LSFD coefficient vector, $\mathbf{b}_{kl}=[ \mathbf{v}_{1k}^{H}\mathbf{h}_{1l},\cdots ,\mathbf{v}_{Mk}^{H}\mathbf{h}_{Ml} ] ^T\in \mathbb{C}^{M}$, and $\mathbf{n}_k=\sum_{m=1}^M{\alpha _{mk}^{*}\mathbf{v}_{mk}^{H}\mathbf{n}_{m}^{\text{ul}}}$, respectively.

\begin{figure*}
\vspace*{-0.7cm}
{{\begin{equation}\tag{15}\label{eq:gama}
\gamma _{k}^{i}=\frac{p_k\mathbf{a}_{k}^{H}\mathbf{b}_{k}^{i}\left( \mathbf{b}_{k}^{i} \right) ^H\mathbf{a}_k}{\mathbf{a}_{k}^{H}\left( \sum_{l=1}^K{p_l\mathbf{\Gamma }_{kl}^{i ,\left( 1 \right)}}+\sum_{l\in \mathcal{P}_k}^{}{p_l\mathbf{\Gamma }_{kl}^{i ,\left( 2 \right)}}-p_k\mathbf{b}_{k}^{i}\left( \mathbf{b}_{k}^{i} \right) ^H+\sigma ^2\mathbf{Z}_{k}^{i} \right) \mathbf{a}_k}.
\end{equation}}
\hrulefill
\vspace*{-0.5cm}
}\end{figure*}

\begin{figure*}
\setcounter{equation}{22}
{{\begin{align}\label{eq:Upsilon_lmmse}\notag
&\left[ \mathbf{\Upsilon }_{kl}^{\left( 1 \right)} \right] _{mm}\!\!=\hat{p}_k\hat{p}_l\tau _{p}^{2}\left[ \left| \text{tr}\left( \left( \mathbf{T}_{mkl\left( 1 \right)}^{H} \right) ^{\frac{1}{2}}\mathbf{R}_{ml}^{\frac{1}{2}} \right) \right| \right. ^2\!\!\!+\!\text{tr}\left( \mathbf{R}_{ml}\mathbf{T}_{mkl\left( 1 \right)} \right) \!+\!\mathbf{\bar{h}}_{ml}^{H}\mathbf{T}_{mkl\left( 1 \right)}^{H}\mathbf{\bar{h}}_{ml}+\mathbf{\bar{h}}_{ml}^{H}\mathbf{S}_{mk}^{H}\mathbf{R}_{ml}\mathbf{S}_{mk}\mathbf{\bar{h}}_{ml}+\!\left| \mathbf{\bar{h}}_{ml}^{H}\mathbf{S}_{mk}\mathbf{\bar{h}}_{ml} \right|^2\\
&\left. +2\text{Re}\left\{ \text{tr}\left( \left( \mathbf{T}_{mkl\left( 1 \right)}^{H} \right) ^{\frac{1}{2}}\mathbf{R}_{ml}^{\frac{1}{2}} \right) \mathbf{\bar{h}}_{ml}^{H}\mathbf{S}_{mk}\mathbf{\bar{h}}_{ml} \right\} \right] +\hat{p}_k\text{tr}\left( \mathbf{R}_{ml}\mathbf{T}_{mkl\left( 2 \right)} \right) +\hat{p}_k\mathbf{\bar{h}}_{ml}^{H}\mathbf{T}_{mkl\left( 2 \right)}^{H}\mathbf{\bar{h}}_{ml}-\left[ \mathbf{\Gamma }_{kl}^{\text{lmmse},\left( 1 \right)} \right] _{mm},
\end{align}}
\vspace*{-0.5cm}
\hrulefill
}\end{figure*}

\section{Spectral Efficiency Analysis}\label{se:SE Analysis}
\vspace{-0.05cm}
In this section, we study the UL SE of CF mMIMO with different estimators and combining schemes. Based on \eqref{eq:sk_LSFD}, an achievable SE of UE $k$ is
\setcounter{equation}{10}
\begin{equation}\label{eq:LSFD_UatF}
\text{SE}_k=\frac{\tau _u}{\tau _c}\log _2\left(1+\gamma _k\right)
\end{equation}
with the effective SINR $\gamma _k$ given by
\begin{equation}\label{eq:LSFD_SINR}
\gamma _k=\frac{p_k\left| \mathbf{a}_{k}^{H}\mathbb{E}\left\{ \mathbf{b}_{kk} \right\} \right|^2}{\mathbf{a}_{k}^{H}\left( \sum_{l=1}^K{p_l\mathbf{\Gamma }_{kl}-p_k\mathbb{E}\left\{ \mathbf{b}_{kk} \right\} \mathbb{E}\left\{ \mathbf{b}_{kk}^{H} \right\} +\sigma ^2\mathbf{Z}_k} \right) \mathbf{a}_k},
\end{equation}
where $\mathbf{\Gamma }_{kl}=[ \mathbb{E}\{\mathbf{v}_{mk}\mathbf{h}_{ml}^{H}\mathbf{v}_{m^{\prime}k}^{H}\mathbf{h}_{m^{\prime}l} \} :\forall m,m^{\prime} ] \in \mathbb{C}^{M\times M}$ and $\mathbf{Z}_k\!=\text{diag}( \mathbb{E}\{ \lVert \mathbf{v}_{1k} \rVert ^2 \} ,\cdots ,\mathbb{E}\{ \lVert \mathbf{v}_{Mk} \rVert ^2 \} ) \in \mathbb{R}^{M\times M}$. And the expectations are with respect to all sources of randomness \cite{8845768}.

Note that we use the use-and-then-forget (UatF) bound as \eqref{eq:LSFD_UatF} which serves a lower bound of the UL ergodic channel capacity of UE $k$ \cite{8187178}. To maximize the effective SINR in \eqref{eq:LSFD_SINR}, $\mathbf{a}_k$ can be optimized by the CPU as
\begin{equation}\label{eq:LSFD_Optimal}
\mathbf{a}_k=\left( \sum_{l=1}^K{p_l\mathbf{\Gamma }_{kl}-p_k\mathbb{E}\left\{ \mathbf{b}_{kk} \right\} \mathbb{E}\left\{ \mathbf{b}_{kk}^{H} \right\} +\sigma ^2\mathbf{Z}_k} \right)^{-1}\!\!\!\mathbb{E}\left\{ \mathbf{b}_{kk} \right\},
\end{equation}
which leads to the maximum SE value
\begin{equation}\label{eq:Max_SE}
\text{SE}_k=\frac{\tau _u}{\tau _c}\log _2\left( 1+p_k\mathbb{E}\left\{ \mathbf{b}_{kk}^{H} \right\} \mathbf{a}_k \right).
\end{equation}
The proof of \eqref{eq:LSFD_Optimal} follows from [15, Lemma B.10] since \eqref{eq:LSFD_Optimal} is a generalized Rayleigh quotient with respect to $\mathbf{a}_{k}$ with a rank-one numerator.

Closed-form SE expressions cannot be obtained when using L-MMSE combining, while Monte Carlo simulations are used to compute the SE with L-MMSE combining. However, we can derive closed-form SE expressions if MR combining adopted. Closed-form SE expressions with different estimators can be similarly formed as $\text{SE}_{k}^{i}=\frac{\tau _u}{\tau _c}\log _2( 1+\gamma _{k}^{i} )$ with $\gamma _{k}^{i}$ shown as \eqref{eq:gama}, where $\mathbf{b}_{k}^{i}=\mathbb{E}\{ \mathbf{b}_{kk} \} \in \mathbb{C}^{M}$ and $\mathbf{Z}_{k}^{i}=\text{diag}( \mathbb{E}\{ \lVert \mathbf{\hat{h}}_{1k}^{i} \rVert ^2 \} ,\cdots ,\mathbb{E}\{ \lVert \mathbf{\hat{h}}_{Mk}^{i} \rVert ^2 \} )$. We define $\mathbf{\Gamma }_{k}^{i}=\sum_{l=1}^K{p_l\mathbf{\Gamma }_{kl}^{i,\left( 1 \right)}}+\sum_{l\in \mathcal{P}_k}^{}{p_l\mathbf{\Gamma }_{kl}^{i,\left( 2 \right)}}-p_k\mathbf{b}_{k}^{i}( \mathbf{b}_{k}^{i} ) ^H+\sigma ^2\mathbf{Z}_{k}^{i}\in \mathbb{C}^{M\times M}$. The SE with maximizing LSFD vector $\mathbf{a}_k=( \mathbf{\Gamma }_{k}^{i} ) ^{-1}\mathbf{b}_{k}^{i}$ is given by
\addtocounter{equation}{1}
\begin{equation}\label{eq:SE_Close}
\text{SE}_{k}^{i}=\frac{\tau _u}{\tau _c}\log _2\left( 1+p_k\left( \mathbf{b}_{k}^{i} \right) ^H\left( \mathbf{\Gamma }_{k}^{i} \right) ^{-1}\mathbf{b}_{k}^{i} \right).
\end{equation}
\subsection{SE with the Phase-Aware MMSE Estimator}\label{se:MMSE_SE}
For MR combining based on the phase-aware MMSE estimator $\mathbf{v}_{mk}=\mathbf{\hat{h}}_{mk}^{\text{mmse}}$, we have $[ \mathbf{Z}_{k}^{\text{mmse}} ] _{mm}=\text{tr}( \hat{p}_k\tau _p\mathbf{\Omega }_{mk} ) +\lVert \mathbf{\bar{h}}_{mk} \rVert ^2$ and $\mathbf{b}_{k}^{\text{mmse}}=\text{diag}( \mathbf{Z}_{k}^{\text{mmse}} )$. And $\mathbf{\Gamma }_{kl}^{\text{mmse},\left( 1 \right)}\in \mathbb{C}^{M\times M}$ is a diagonal matrix with the $\left( m,m \right)$-th element given by
\begin{align}\notag
\left[ \mathbf{\Gamma }_{kl}^{\text{mmse},\left( 1 \right)} \right] _{mm}&=\hat{p}_k\tau _p\text{tr}\left( \mathbf{R}_{ml}\mathbf{\Omega }_{mk} \right) +\mathbf{\bar{h}}_{mk}^{H}\mathbf{R}_{ml}\mathbf{\bar{h}}_{mk}\\
&+\hat{p}_k\tau _p\mathbf{\bar{h}}_{ml}^{H}\mathbf{\Omega }_{mk}\mathbf{\bar{h}}_{ml}+\left| \mathbf{\bar{h}}_{mk}^{H}\mathbf{\bar{h}}_{ml} \right|^2.
\end{align}
The computation of above results follow similar steps as \cite{8809413} and \cite{8620255}. Moreover,
\begin{equation}
\mathbf{\Gamma }_{kl}^{\text{mmse},\left( 2 \right)}=\begin{cases}
	\hat{p}_k\hat{p}_l\tau _{p}^{2}\mathbf{z}_{kl}^{\text{mmse}}\left( \mathbf{z}_{kl}^{\text{mmse}} \right) ^H, &l\in \mathcal{P}_k\backslash \{k\}\\
	\mathbf{b}_{k}^{\text{mmse}}\left( \mathbf{b}_{k}^{\text{mmse}} \right) ^H-\mathbf{L}_{k}^{2},&l=k\\
\end{cases}
\end{equation}
where $\mathbf{z}_{kl}^{\text{mmse}}=[ \text{tr}( \mathbf{R}_{1l}\mathbf{\Psi }_{1k}^{-1}\mathbf{R}_{1k} ) ,\cdots ,\text{tr}( \mathbf{R}_{Ml}\mathbf{\Psi }_{Mk}^{-1}\mathbf{R}_{Mk} ) ] ^T$ and $\mathbf{L}_k=\text{diag}( \lVert \mathbf{\bar{h}}_{1k} \rVert ^2,\cdots ,\lVert \mathbf{\bar{h}}_{Mk} \rVert ^2 ) $.
So $\mathbf{\Gamma }_{k}^{\text{mmse}}=\sum_{l=1}^K{p_l\mathbf{\Gamma }_{kl}^{\text{mmse},\left( 1 \right)}}+\sum_{l\in \mathcal{P}_k}^{}{p_l\mathbf{\Gamma }_{kl}^{\text{mmse},( 2 )}}-p_k\mathbf{b}_{k}^{\text{mmse}}( \mathbf{b}_{k}^{\text{mmse}} ) ^H+\sigma ^2\mathbf{Z}_{k}^{\text{mmse}}$. So we can derive the closed-form SE based on the phase-aware MMSE estimator from \eqref{eq:SE_Close} using the matrices and vectors that we have computed.
\subsection{SE with the Phase-Aware EW-MMSE Estimator}\label{se:EW_MMSE_SE}
If MR combining based on the phase-aware EW-MMSE estimator $\mathbf{v}_{mk}=\mathbf{\hat{h}}_{mk}^{\text{ew}}$ is adopted, $[ \mathbf{Z}_{k}^{\text{ew}} ] _{mm}=\text{tr}( \mathbf{\Sigma }_{mk}) +\lVert \mathbf{\bar{h}}_{mk} \rVert ^2$ and $\mathbf{b}_{k}^{\text{ew}}=[ \hat{p}_k\tau _p\text{tr}( \mathbf{D}_{mk}\mathbf{\Lambda }_{mk}^{-1}\mathbf{D}_{mk} ) +\lVert \mathbf{\bar{h}}_{mk} \rVert ^2:m=1,\cdots ,M ] ^T\!\!\!\in \mathbb{R}^M$. Besides, $\mathbf{\Gamma }_{kl}^{\text{ew},\left( 1 \right)}\in \mathbb{C}^{M\times M}$ is a diagonal matrix with the $\left( m,m \right)$-th element given by
\begin{align}\notag
\left[ \mathbf{\Gamma }_{kl}^{\text{ew},\left( 1 \right)} \right] _{mm}&=\text{tr}\left( \mathbf{R}_{ml}\mathbf{\Sigma }_{mk} \right) +\mathbf{\bar{h}}_{mk}^{H}\mathbf{R}_{ml}\mathbf{\bar{h}}_{mk}\\
&+\mathbf{\bar{h}}_{ml}^{H}\mathbf{\Sigma }_{mk}\mathbf{\bar{h}}_{ml}+\left| \mathbf{\bar{h}}_{mk}^{H}\mathbf{\bar{h}}_{ml} \right|^2.
\end{align}
\begin{equation}
\mathbf{\Gamma }_{kl}^{\text{ew},\left( 2 \right)}=\begin{cases}
	\hat{p}_k\hat{p}_l\tau _{p}^{2}\mathbf{z}_{kl}^{\text{ew}}\left( \mathbf{z}_{kl}^{\text{ew}} \right) ^H, &l\in \mathcal{P}_k\backslash \{k\}\\
	\mathbf{b}_{k}^{\text{ew}}\left( \mathbf{b}_{k}^{\text{ew}} \right) ^H-\mathbf{L}_{k}^{2},&l=k\\
\end{cases}
\end{equation}
where $\mathbf{z}_{kl}^{\text{ew}}=[ \text{tr}( \mathbf{D}_{1l}\mathbf{\Lambda }_{1k}^{-1}\mathbf{D}_{1k} ) ,\cdots ,\text{tr}( \mathbf{D}_{Ml}\mathbf{\Lambda }_{Mk}^{-1}\mathbf{D}_{Mk} )] ^T$. So $\mathbf{\Gamma }_{k}^{\text{ew}}=\sum_{l=1}^K{p_l\mathbf{\Gamma }_{kl}^{\text{ew},\left( 1 \right)}}+\sum_{l\in \mathcal{P}_k}^{}{p_l\mathbf{\Gamma }_{kl}^{\text{ew},\left( 2 \right)}}-p_k\mathbf{b}_{k}^{\text{ew}}( \mathbf{b}_{k}^{\text{ew}}) ^H+\sigma ^2\mathbf{Z}_{k}^{\text{ew}}$. We can derive the closed-form SE expression of the phase-aware EW-MMSE estimator from \eqref{eq:SE_Close} using the matrices and vectors that we have computed.
\subsection{SE with the LMMSE Estimator}\label{se:LMMSE_SE}
If we use MR combining based on the LMMSE estimator $\mathbf{v}_{mk}=\mathbf{\hat{h}}_{mk}^{\text{lmmse}}$, we have $[ \mathbf{Z}_{k}^{\text{lmmse}} ] _{mm}=\hat{p}_k\tau _p\text{tr}( \mathbf{\Omega }'_{mk} )$ and $\mathbf{b}_{k}^{\text{lmmse}}=\text{diag}( \mathbf{Z}_{k}^{\text{lmmse}} )$. Moreover, $\mathbf{\Gamma }_{kl}^{\text{lmmse},\left( 1 \right)}\in \mathbb{C}^{M\times M}$ is a diagonal matrix with the $\left( m,m \right)$-th element given by
\begin{equation}
\left[ \mathbf{\Gamma }_{kl}^{\text{lmmse},\left( 1 \right)} \right] _{mm}=\hat{p}_k\tau _p\left[ \text{tr}\left( \mathbf{R}_{ml}\mathbf{\Omega }'_{mk} \right) +\mathbf{\bar{h}}_{ml}^{H}\mathbf{\Omega }'_{mk}\mathbf{\bar{h}}_{ml} \right].
\end{equation}
Besides, we can obtain $\mathbf{\Gamma }_{kl}^{\text{lmmse},\left( 2 \right)}$ as
\begin{equation}
\mathbf{\Gamma }_{kl}^{\text{lmmse},\left( 2 \right)}=\mathbf{\Upsilon }_{kl}^{\left( 1 \right)}+\mathbf{d}_{kl}^{\text{lmmse}}\left( \mathbf{d}_{kl}^{\text{lmmse}} \right) ^H-\mathbf{\Upsilon }_{kl}^{\left( 2 \right)},
\end{equation}
 where $\mathbf{\Upsilon }_{kl}^{\left( 1 \right)}\in \mathbb{C}^{M\times M}$, $\mathbf{\Upsilon }_{kl}^{\left( 2 \right)}\in \mathbb{C}^{M\times M}$ are diagonal matrices and $\mathbf{d}_{kl}^{\text{lmmse}}=\text{diag}\{ ( \mathbf{\Upsilon }_{kl}^{\left( 2 \right)} ) ^{\frac{1}{2}} \}$. The $\left( m,m \right) $-th element of $\mathbf{\Upsilon }_{kl}^{\left( 1 \right)}$ is given by \eqref{eq:Upsilon_lmmse}, where $\mathbf{S}_{mk}=\mathbf{R}'_{mk}(\mathbf{\Psi }'_{mk})^{-1}, \mathbf{T}_{mkl\left(1\right)}=\mathbf{S}_{mk}\mathbf{R}_{ml}\mathbf{S}_{mk}^{H}$ and $\mathbf{T}_{mkl\left( 2 \right)}=\tau _p\mathbf{S}_{mk}\mathbf{\Psi }'_{mk}\mathbf{S}_{mk}^{H}-\hat{p}_l\tau _{p}^{2}\mathbf{S}_{mk}\mathbf{R}'_{ml}\mathbf{S}_{mk}^{H}$, respectively. And the $( m,m) $-th element of $\mathbf{\Upsilon }_{kl}^{\left( 2 \right)}$ is
\addtocounter{equation}{1}
\begin{equation}
\left[ \mathbf{\Upsilon }_{kl}^{\left( 2 \right)} \right] _{mm}=\hat{p}_k\hat{p}_l\tau _{p}^{2}\text{tr}\left( \mathbf{R}'_{ml}\mathbf{R}'_{mk}\left( \mathbf{\Psi }'_{mk} \right) ^{-1} \right) ^2.
\end{equation}
So $\mathbf{\Gamma }_{k}^{\text{lmmse}}=\sum_{l=1}^K{p_l\mathbf{\Gamma }_{kl}^{\text{lmmse},\left( 1 \right)}}+\sum_{l\in \mathcal{P}_k}^{}{p_l\mathbf{\Gamma }_{kl}^{\text{lmmse},\left( 2 \right)}}-p_k\mathbf{b}_{k}^{\text{lmmse}}( \mathbf{b}_{k}^{\text{lmmse}} ) ^H+\sigma ^2\mathbf{Z}_{k}^{\text{lmmse}}$ and we can obtain the closed-form SE expression based on the LMMSE estimator from \eqref{eq:SE_Close}. Note that the phase-aware MMSE estimator achieves better SE than other estimators since it makes use of prior phase knowledge, which will be demonstrated in Section IV.
\section{Numerical Results}\label{se:Numerical Results}
We consider APs and UEs are uniformly distributed in a $1\times1\,\text{km}^2$ area with a wrap-around scheme \cite{8187178}. All AP-UE pairs have LoS paths and the pathloss is computed by the COST 321 Walfish-Ikegami model as
\begin{equation}
\beta _{mk}\left[ \text{dB} \right] =-30.18-26\log _{10}\left( \frac{d_{mk}}{1\,\text{m}} \right) +F_{mk},
\end{equation}
where $d_{mk}$ is the distance between AP $m$ and UE $k$ (taking an $11\,\text{m}$ height difference into account). The Rician $\kappa$-factor is computed as $\kappa _{mk}=10^{1.3-0.003d_{mk}}$. We model the shadow fading $F_{mk}$ as in \cite{7827017} with $F_{mk}=\sqrt{\delta _f}a_m+\sqrt{1-\delta _f}b_k$, where $a_m\sim\mathcal{N}(0,\delta _{\text{sf}}^{2} )$ and $b_k\sim \mathcal{N}(0,\delta _{\text{sf}}^{2} )$ are independent random variables and $\delta _f$ is the shadow fading parameter. The covariance functions of $a_m$ and $b_k$ are $\mathbb{E}\{ a_ma_{m'} \} =2^{-\frac{d_{mm'}}{d_{\text{dc}}}}$, $\mathbb{E}\{ b_kb_{k'} \} =2^{-\frac{d_{kk'}}{d_{\text{dc}}}}$ where $d_{mm'}$ and $d_{kk'}$ are the geographical distances between AP $m$-AP $m'$ and UE $k$-UE $k'$, respectively, $d_{\text{dc}}$ is the decorrelation distance depending on the environment. Let $\delta _f=0.5$, $d_{\text{dc}}=100\,\text{m}$ and $\delta _{\text{sf}}=8$ in this paper. The large-scale coefficients of $\mathbf{h}_{mk}$ are given by
\begin{equation}
\beta _{mk}^{\text{LoS}}=\frac{\kappa _{mk}}{\kappa _{mk}+1}\beta _{mk},\ \beta _{mk}^{\text{NLoS}}=\frac{1}{\kappa _{mk}+1}\beta _{mk}.
\end{equation}
Each AP is equipped with a uniform linear array (ULA) with omnidirectional antennas so the $n$-th element of the deterministic LoS component $\mathbf{\bar{h}}_{m,k}\in \mathbb{C}^N$ can be written as $\left[ \mathbf{\bar{h}}_{mk} \right] _n=\sqrt{\beta _{mk}^{\text{LoS}}}e^{j2\pi d_H\left( n-1 \right) \sin \left( \theta _{mk} \right)}$, where $\theta _{mk}$ is the angle of arrival to the UE $k$ seen from AP $m$ and $d_H$ denotes the antenna spacing parameter (in fractions of the wavelength). The spatial correlation matrix $\mathbf{R}_{mk}$ is generated based on the Gaussian local scattering model \cite{8187178}. The $\left( l,n \right)$-th element of $\mathbf{R}_{mk}$ is given by
\begin{equation}
\left[ \mathbf{R}_{mk} \right] _{ln}=\frac{\beta _{mk}^{\text{NLoS}}}{\sqrt{2\pi}\sigma _{\varphi}}\int_{-\infty}^{+\infty}{e^{j2\pi d_H\left( l-n \right) \sin \left( \theta _{mk}+\delta \right)}e^{-\frac{\delta ^2}{2\sigma _{\varphi}^{2}}}d\delta},
\end{equation}
where $\delta \sim \mathcal{N}( 0,\sigma _{\varphi}^{2})$ is a Gaussian distributed deviation from $\theta _{mk}$ with angular standard deviation (ASD) $\sigma _{\varphi}$. All the UEs transmit with power $200\,\text{mW}$, the bandwidth is $20\,\text{MHz}$, the noise power $\sigma ^2=-94\,\text{dBm}$, and every coherence block contains $\tau _c=200$ channel uses where $\tau _p=10$ channel uses are reserved for pilot transmission.

Figure 1 shows the UL SE averaged over random UE locations and shadow fading realizations as a function of the number of APs $M$ for different $N$ with MR/L-MMSE combining based on the MMSE estimator. The average SE grows with $N$, e.g., $92.02\%$ improvement with MR combining for $N=4,M=100$ compared with the $N=1,M=100$ scenario. Moreover, L-MMSE combining performs much better than MR combining, e.g., $27.78\%$ SE improvement for L-MMSE combining compared with that of MR combining for $N=4,M=100$. The performance gap between L-MMSE and MR combining becomes larger with the increase of $N$ since L-MMSE combining can use all antennas on each AP to suppress interference, which means that L-MMSE combining should be advocated in the scenario with multi-antenna APs.
\begin{figure}[t]
\setlength{\abovecaptionskip}{-0.15cm}
\vspace{-0.2cm}
\centering
\includegraphics[scale=0.5]{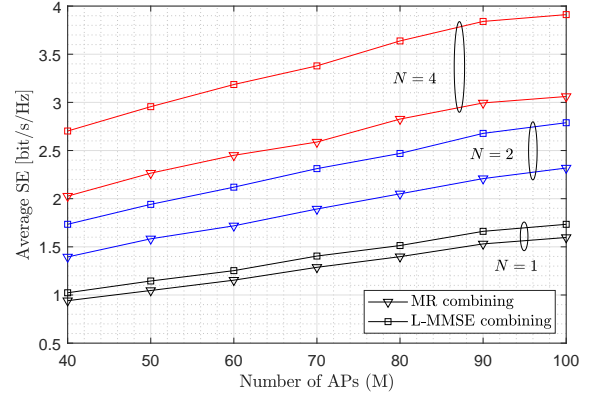}
\caption{Average SE against the number of APs $M$ with different combining schemes for $K=40,N=\left[ 1,2,4 \right]$ and $\sigma _{\varphi}=15^{\circ}$.
\label{fig1:avSE}}
\vspace{-0.7cm}
\end{figure}

Figure 2 shows the cumulative distribution function (CDF) curves for the SE per UE over spatially correlated/uncorrelated Rician fading channels with MR/L-MMSE combining based on the MMSE estimator. The spatial channel correlation increases as $\sigma _{\varphi}$ reduces. Let $\sigma _{\varphi}=5^{\circ}/30^{\circ}$ represent strong/moderate spatial correlation, respectively, and $\mathbf{R}_{mk}=\beta _{mk}^{\text{NLoS}}\mathbf{I}_N$ is diagonal in the uncorrelated fading scenario. Note that the SE benefits from the spatial correlation since the spatial channel correlation improves the level of favorable propagation and channel estimation quality. This finding coincides with the one in \cite[Sec. 4.1]{8187178}. The CDF curve in moderate spatial correlation scenario approximately coincides with that of the uncorrelated fading scenario. Besides, the strong spatial correlation outperforms the moderate spatial correlation and the gap at $95\%$ likely SE points between the strong spatial correlation and the moderate spatial correlation are $11.64\%$ and $29.78\%$ for MR/L-MMSE combining, respectively.

\begin{figure}[t]
\setlength{\abovecaptionskip}{-0.1cm}
\vspace{-0.1cm}
\centering
\includegraphics[scale=0.5]{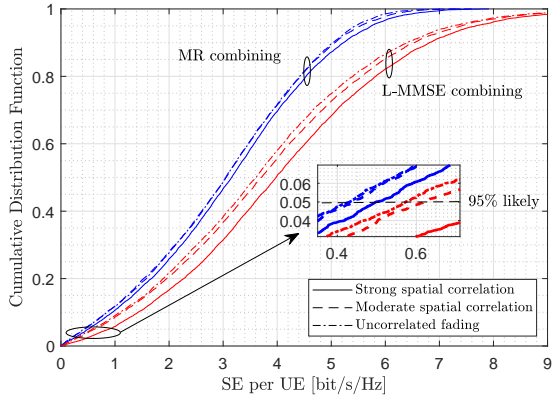}
\caption{ CDF of SE per UE for $M=100,K=40$ and $N=4$ over correlated/uncorrelated Rician fading.
\label{fig2:corre}}
\vspace{-0.3cm}
\end{figure}

\begin{figure}[t]
\setlength{\abovecaptionskip}{-0.1cm}
\centering
\includegraphics[scale=0.5]{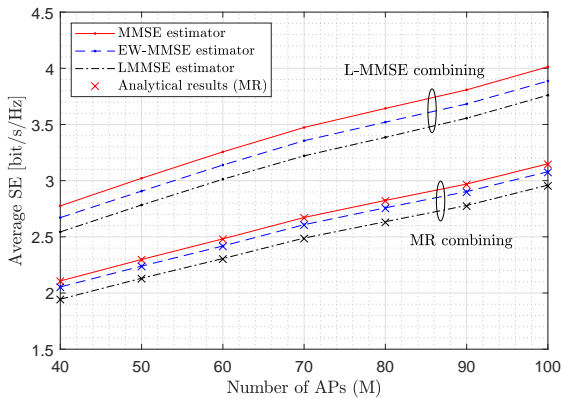}
\caption{Average SE against the number of APs $M$ with different estimators and combining schemes for $K=40, N=4$ and $\sigma _{\varphi}=15^{\circ}$.
\label{fig3:estimator}}
\vspace{-0.6cm}
\end{figure}

Figure 3 shows the average SE as a function of the number of APs $M$ for $N=4$ with MR/L-MMSE combining based on different estimators. For MR combining, markers ``$\times$'' generated by analytical results from \eqref{eq:SE_Close} overlap with the curves generated by simulations, respectively, which validates our derived closed-form SE expressions. In the considered scenarios, the MMSE estimator with all prior information undoubtedly achieves the best performance. Moreover, the EW-MMSE estimator with phase-shifts and partial large-scale fading knowledge outperforms the LMMSE estimator with only large-scale fading knowledge but no phase-shifts knowledge. The performance gap between the EW-MMSE estimator and the LMMSE estimator are $3.88\%$ and $3.24\%$ for MR/L-MMSE combining, respectively, for $M=100$. Furthermore, we observe that performance gap due to the lack of phase-shifts knowledge between the MMSE estimator and the LMMSE estimator are $6.04\%$ and $6.30\%$ for MR/L-MMSE combining, respectively, for $M=100$. Note that the phase-shift is significant for the SE of CF mMIMO systems so it is worth acquiring phase-shifts to use a more advanced estimator.

\section{Conclusions}\label{se:conclusion}
In this paper, we study the UL SE of a CF mMIMO system over spatially correlated Rician fading channels, where the phase-shift of the LoS component is modeled randomly. For phase-aware MMSE, phase-aware EW-MMSE, and LMMSE estimators, we derive UL SE expressions for any combining scheme and compute closed-form SE expressions for MR combining. It is important to find that L-MMSE combining performs much better than MR combining in multi-antenna APs scenarios. Moreover, the SE grows with the number of antennas per AP and benefits from the spatial correlation. Finally, the MMSE estimator achieves the best performance and the SE of the LMMSE estimator is lower than the one of other estimators due to the lack of phase-shift knowledge. In the future work, we will consider multi-antenna UEs with uplink precoding, power control, and algorithmic scalability to enable implementation of large CF mMIMO networks.

\bibliographystyle{IEEEtran}
\bibliography{IEEEabrv,Ref}

\begin{thebibliography}{10}
\providecommand{\url}[1]{#1}
\csname url@samestyle\endcsname
\providecommand{\newblock}{\relax}
\providecommand{\bibinfo}[2]{#2}
\providecommand{\BIBentrySTDinterwordspacing}{\spaceskip=0pt\relax}
\providecommand{\BIBentryALTinterwordstretchfactor}{4}
\providecommand{\BIBentryALTinterwordspacing}{\spaceskip=\fontdimen2\font plus
\BIBentryALTinterwordstretchfactor\fontdimen3\font minus
  \fontdimen4\font\relax}
\providecommand{\BIBforeignlanguage}[2]{{%
\expandafter\ifx\csname l@#1\endcsname\relax
\typeout{** WARNING: IEEEtran.bst: No hyphenation pattern has been}%
\typeout{** loaded for the language `#1'. Using the pattern for}%
\typeout{** the default language instead.}%
\else
\language=\csname l@#1\endcsname
\fi
#2}}
\providecommand{\BIBdecl}{\relax}
\BIBdecl

\bibitem{7827017}
H.~Q. Ngo, A.~Ashikhmin, H.~Yang, E.~G. Larsson, and T.~L. Marzetta,
  ``{Cell-free massive {MIMO} versus small cells},'' \emph{IEEE Trans. Wireless
  Commun.}, vol.~16, no.~3, pp. 1834--1850, Mar. 2017.

\bibitem{zhang2019multiple}
J.~Zhang, E.~Bj{\"o}rnson, M.~Matthaiou, D.~W.~K. Ng, H.~Yang, and D.~J. Love,
  ``Prospective multiple antenna technologies for beyond {5G},'' \emph{IEEE J.
  Sel. Areas Commun.}, vol.~38, no.~8, pp. 1637--1660, Aug. 2020.

\bibitem{7917284}
E.~{Nayebi}, A.~{Ashikhmin}, T.~L. {Marzetta}, H.~{Yang}, and B.~D. {Rao},
  ``Precoding and power optimization in cell-free massive {MIMO} systems,''
  \emph{IEEE Trans. Wireless Commun.}, vol.~16, no.~7, pp. 4445--4459, Jul.
  2017.

\bibitem{8901451}
S.~{Buzzi}, C.~{D'Andrea}, A.~{Zappone}, and C.~{D'Elia}, ``User-centric {5G}
  cellular networks: {R}esource allocation and comparison with the cell-free
  massive {MIMO} approach,'' \emph{IEEE Trans. Wireless Commun.}, vol.~19,
  no.~2, pp. 1250--1264, Feb. 2020.

\bibitem{8943119}
D.~{Wang}, M.~{Wang}, P.~{Zhu}, J.~{Li}, J.~{Wang}, and X.~{You}, ``Performance
  of network-assisted full-duplex for cell-free massive {MIMO},'' \emph{IEEE
  Trans. Commun.}, vol.~68, no.~3, pp. 1464--1478, Mar. 2020.

\bibitem{8845768}
E.~{Bj{\"o}rnson} and L.~{Sanguinetti}, ``Making cell-free massive {MIMO}
  competitive with {MMSE} processing and centralized implementation,''
  \emph{IEEE Trans. Wireless Commun.}, vol.~19, no.~1, pp. 77--90, Jan. 2020.

\bibitem{9004558}
J.~{Zheng}, J.~{Zhang}, L.~{Zhang}, X.~{Zhang}, and B.~{Ai}, ``Efficient
  receiver design for uplink cell-free massive {MIMO} with hardware
  impairments,'' \emph{IEEE Trans. Veh. Technol.}, vol.~69, no.~4, pp.
  4537--4541, Apr. 2020.

\bibitem{8891922}
H.~{Masoumi} and M.~J. {Emadi}, ``Performance analysis of cell-free massive
  {MIMO} system with limited fronthaul capacity and hardware impairments,''
  \emph{IEEE Trans. Wireless Commun.}, vol.~19, no.~2, pp. 1038--1053, Feb.
  2020.

\bibitem{8886730}
P.~{Liu}, K.~{Luo}, D.~{Chen}, and T.~{Jiang}, ``Spectral efficiency analysis
  of cell-free massive {MIMO} systems with zero-forcing detector,'' \emph{IEEE
  Trans. Wireless Commun.}, vol.~19, no.~2, pp. 795--807, Feb. 2020.

\bibitem{8869794}
Y.~{Zhang}, M.~{Zhou}, H.~{Cao}, L.~{Yang}, and H.~{Zhu}, ``On the performance
  of cell-free massive {MIMO} with mixed-{ADC} under {R}ician fading
  channels,'' \emph{IEEE Commun. Lett.}, vol.~24, no.~1, pp. 43--47, Jan. 2020.

\bibitem{8809413}
{\"O}.~{{\"O}zdogan}, E.~{Bj{\"o}rnson}, and J.~{Zhang}, ``Performance of
  cell-free massive {MIMO} with {R}ician fading and phase shifts,'' \emph{IEEE
  Trans. Wireless Commun.}, vol.~18, no.~11, pp. 5299--5315, Nov. 2019.

\bibitem{133}
{\"O}.~T. {Demir} and E.~Bj{\"o}rnson, ``Max-min fair wireless-powered
  cell-free massive {MIMO} for uncorrelated {R}ician fading channels,''
  \emph{arXiv:2001.09296}, 2020.

\bibitem{8620255}
{\"O}.~{{\"O}zdogan}, E.~{Bj{\"o}rnson}, and E.~G. {Larsson}, ``Massive {MIMO}
  with spatially correlated {R}ician fading channels,'' \emph{IEEE Trans.
  Commun.}, vol.~67, no.~5, pp. 3234--3250, May 2019.

\bibitem{9053886}
{\"O}.~T. {Demir} and E.~{Bj{\"o}rnson}, ``Large-scale fading precoding for
  maximizing the product of {SINR}s,'' in \emph{Proc. IEEE ICASSP}, May 2020.

\bibitem{8187178}
E.~Bj{\"o}rnson, J.~Hoydis, and L.~Sanguinetti, ``{Massive MIMO networks:
  Spectral, energy, and hardware efficiency},'' \emph{Foundations and
  Trends{\textregistered} in Signal Processing}, vol.~11, no. 3-4, pp.
  154--655, 2017.

\end{thebibliography}
\end{document}